\documentclass[12pt]{article}

\usepackage{amsmath, amssymb, slashed}
\usepackage[dvips]{graphicx}
\usepackage{epsf}
\usepackage{color}

\begin{document}

\begin{titlepage}
\begin{center}

\hfill IPMU-12-0142\\
\hfill ICRR-report-622-2012-11\\
\hfill \today

\vspace{1.5cm}
{\large\bf Pure Gravity Mediation of Supersymmetry Breaking at the LHC}
\vspace{2.0cm}

{\bf Biplob Bhattacherjee}$^{(a)}$,
{\bf Brian Feldstein}$^{(a)}$,
{\bf Masahiro Ibe}$^{(b,a)}$,
\\
{\bf Shigeki Matsumoto}$^{(a)}$
and
{\bf Tsutomu T. Yanagida}$^{(a)}$

\vspace{1.0cm}
{\it
$^{(a)}${Kavli IPMU, University of Tokyo, Kashiwa, 277-8583, Japan} \\
$^{(b)}${\it ICRR, University of Tokyo, Kashiwa, 277-8583, Japan }
}
\vspace{2.0cm}

\abstract{
Supersymmetric theories which can allow for a $125$\,GeV Higgs mass and also solve the naturalness and susy flavor problems now require a fair degree of complexity.  
Here we consider the simplest possibility for supersymmetry near the weak scale, 
but with the requirement of naturalness dropped. 
In ``pure gravity mediation'', all supersymmetric particles except for the gauginos lie at tens to thousands
 of TeV, with the gauginos obtaining loop suppressed masses automatically 
by anomaly mediation and higgsino threshold corrections.   
The gauginos are the lightest superpartners, and we investigate 
the current collider constraints on their masses, as well as the future reach of the LHC. 
We consider gluino pair production with a jets + missing energy signature, 
as well as events with disappearing charged tracks caused by charged winos decaying 
into their neutral partners.  
We show that presently, gluino masses less than about $1$\,TeV and wino masses less than 
about $300$\,GeV are excluded, and that the $14$\,TeV LHC can probe gluino masses up to 
about $2$\,TeV and wino masses up to $1$\,TeV.
}

\end{center}
\end{titlepage}
\setcounter{footnote}{0}

\section{Introduction}
\label{sec: introduction}
With data from the Large Hadron Collider (LHC) 
continuing to show no evidence for supersymmetry (SUSY), 
and with a Higgs-like particle with a mass of  $\sim 125$GeV  
having now been observed, the situation for natural electroweak symmetry breaking 
has grown increasingly severe.  
Natural weak scale supersymmetry now faces a set of difficult challenges:
\begin{itemize}
\item{In the minimal supersymmetric standard model (MSSM), 
the Higgs mass is required to be less than the $Z$-boson mass at tree level.  
While loop corrections increase this upper bound, their logarithmic nature implies that, for a Higgs mass of $125$\,GeV, the superpartner masses should be at least $10$-$100$\,TeV \cite{Higgs,TU-363}.  
Such heavy superpartners would seem to be at odds with naturalness, though non-minimal implementations of weak scale SUSY with, for example, a new singlet field, or new gauge interactions, can potentially raise the Higgs mass with some cost in complexity.}
\item{Even if a mechanism is introduced to raise the Higgs mass, 
one must explain the absence of any observation of superpartners. 
While certain forms of  $R$-parity violation may be introduced to weaken the current collider 
limits\,\cite{R1, R2, R3}, 
this removes the dark matter candidate from the theory, and again adds some degree of complexity.}
\item{Flavor changing neutral current (FCNC) constraints on the superpartner mass matrices are severe.  With generic mass matrices and ${\cal O}(1)$ CP violating phases, $K^0-\bar{K}^0$ 
mixing would seem to require the squark masses to be greater than perhaps $1000$ TeV, 
precluding a natural weak scale\,\cite{Gabbiani:1996hi}.
In this case again, it is possible to introduce an additional assumption into the theory, 
such as gauge mediation\,\cite{GMSB}, or dilaton mediation\,\cite{D1, D2},
but once more the theory requires additional complexity.
}
\item{In the simplest SUSY breaking mediation mechanism, gravity mediation\,\cite{SUGRA},
even if one assumes flavor universality of the superpartner masses in order to suppress FCNCs, 
one is still faced with a serious difficulty. 
In order to obtain gaugino masses as large as those of the other superpartners
(as demanded by phenomenology if supersymmetry is present at the weak scale), 
we require gauginos to obtain masses at order ${1/M_{\rm PL}}$.  
This in turn requires the presence of an $F$-term supersymmetry breaking vacuum expectation value associated with a singlet chiral superfield  known as the Polonyi field.  
Since the origin for this singlet field is not an enhanced point of any symmetry, generically during inflation the field will be located very far from the origin.  
The Polonyi field typically has a mass of the order of the gravitino mass, $m_{3/2}$, 
and  a long lifetime ${\cal O}(M_{\rm PL}^2/m^3_{3/2})$.
After the expansion rate of the universe drops below $m_{3/2}$ the coherent oscillations of this field thus 
dominate the energy density of the universe. 
If supersymmetry is present at the weak scale, then the Polonyi field typically decays 
after big bang nucleosynthesis (BBN), and is phenomenologically excluded\,\cite{Polonyi}.
}
\end{itemize}

In spite of these issues, there remains a number of striking pieces of circumstantial evidence 
in favor of supersymmetry, independent of naturalness.  
Gauge coupling unification in the presence of Standard Model superpartners 
is successful at the few percent precision level. 
Sufficient proton stability as well as a stable dark matter candidate may be simultaneously 
obtained through $R$-parity conservation.  
In addition, string theory- a leading candidate for a quantum theory of gravity- is thought to require supersymmetry, with supersymmetry breaking to occur at some unknown scale below the Planck scale.  
In fact, if one simply relax the requirement on naturalness,
then all of the other benefits of supersymmetry may be maintained, 
without disturbing the simplicity and minimality of the supersymmetric sector.

Without naturalness, the situation for supersymmetry becomes the following:
\begin{itemize}
\item{No new mechanism is required to raise the Higgs mass, as this feat is accomplished by heavy superpartner masses alone.}
\item{Superpartners have not yet been detected since their masses have left them out of reach of all searches performed to date.}
\item{FCNC processes can be sufficiently suppressed due to the heavy sfermions, and gravity mediation in its simplest form- without any ad hoc assumptions concerning flavor universality- may be used to communicate supersymmetry breaking to the visible sector.}
\item{The wino is the lightest superpartner, and is a stable dark matter candidate.  The appropriate relic abundance may be obtained either through thermal freeze-out, or non-thermally through gravitino decays, depending on the wino mass.}
\item{No Polonyi field is required, since the gauginos may now obtain masses smaller than the other superpartners without causing difficulty for phenomenology.  Gauginos obtain masses automatically through anomaly mediation and higgsino threshold corrections, suppressed compared to the other superpartners by factors of order 
$10^{-3}-10^{-2}$\,\cite{Giudice:1998xp,Randall:1998uk,hep-ph/9205227}.  
No new assumptions or model building is required to obtain these masses.}
\end{itemize}

There is only one potentially serious difficulty faced by this scenario:  It may be out of reach of currently planned experimental efforts.  Indeed, $CP$ violation in $K^0-\bar{K}^0$ mixing constrains the $1-2$ elements of the left and right down squark mass matrices as follows\,\cite{Gabbiani:1996hi}:
\begin{eqnarray}
\sqrt{\tilde{m}_{LL}\tilde{m}_{RR}}
&\gtrsim& 
4000\,
{\rm TeV}\times\sqrt{\left|{\rm Im}\left(\frac{m^{d\,\,2}_{12,LL}}{\tilde{m}^2_{LL}}\frac{m^{d\,\,2}_{12,RR}}
{\tilde{m}^2_{RR}}\right)\right|} \ ,
\label{eq:d1}\\
 \tilde{m}_{LL}
&\gtrsim&
700\,{\rm TeV}\times\sqrt{\left|{\rm Im}\left(\frac{m^{d\,\,2}_{12,LL}}{\tilde{m}^2_{LL}}\right)^2 \right|} 
 \label{eq:d2}\ ,
\end{eqnarray}
where $\tilde{m}_A={\cal O}(m_{3/2})$ is the average of the  first and second generation squark masses of type A.
The naive limit on the sfermion mass scale is thus roughly a few thousand TeV, i.e. $m_{3/2} \gtrsim 
{\cal O}(10^3)$\,TeV. 
Remembering that the anomaly mediated gaugino masses are of order $10^{-3}-10^{-2} m_{3/2}$, this leads to gauginos with masses which might be expected to be larger than $\sim$ TeV, and detection of superpartners at the LHC will be challenging.  This issue is exacerbated by the fact that the correct thermal relic abundance of wino dark matter 
is obtained for wino masses of about $2.7$\,TeV~\cite{hep-ph/0610249}.  Note that for such large wino masses, dark 
matter direct and indirect detection will unfortunately be out of reach for the foreseeable future.

In spite of this somewhat pessimistic argument, there are good reasons to be hopeful.  
As will be discussed in detail later in this paper, detection of the wino and gluino at the LHC require roughly $m_{\rm wino} \lesssim$1\,TeV and $m_{\rm gluino} \lesssim 2$\,TeV respectively.  
There is thus some tension here with the bounds on these masses suggested by $K^0-\bar{K}^0$ 
mixing, but this tension is clearly very mild. 
Even minor order 1 suppressions of the off-diagonal down squark mass matrix elements, or mildly suppressed $CP$ violating phases, etc, would be sufficient to allow LHC discovery of at least one of these superpartners.\footnote{It is also worth noting that if the squarks and quarks share in any flavor symmetries, then this would also generically lead to suppressions in $K^0-\bar{K}^0$ mixing, and lower the allowed superparticle masses.}

Note that wino masses smaller than the thermal relic value may still yield appropriate relic densities through production in gravitino decays, as long as the reheating temperature $T_R$ is set appropriately.  
Moreover, if $T_R$ is higher than about $2 \times 10^9$ GeV, as required by models of thermal leptogenesis \cite{leptogenesis}, then the wino mass \emph{must} be smaller than the thermal relic value in order to avoid overclosing the universe.  The essential point is that, as the reheating temperature is increased, the gravitino number density, and therefore the resulting wino number density, go up.  This requires decreasing the wino mass in order to keep the energy density fixed, and thus with thermal leptogenesis one obtains an upper bound $M_2 \lesssim1 $\,TeV \cite{Ibe:2011aa}.\footnote{It is worth noting that many models of inflation also require high reheating temperatures, and will thus similarly require sufficiently small wino masses.}  Such values of the wino mass may be obtained with susy partner scales easily high enough to accommodate a $125$\,GeV Higgs boson. 
Thus while this scenario, dubbed ``pure gravity mediation" (PGM) in \cite{Ibe:2011aa, Ibe:2012hu},  leaves little guarantee that superpartners will be discovered 
in the near future, the prospects are reasonably good.  
In this paper, we will analyze the detection prospects of this model at the LHC in detail. 

After reviewing details of the model in section 2, we will turn in section 3.1 to summarizing the current bounds on the gaugino masses coming from LHC searches.  
The first signal we will consider involves the direct production of gluino pairs, with the gluinos decaying into quarks and  winos and leading to a standard jets + missing energy signature.  
The resulting limits depend somewhat on the squark mass spectrum, which determines whether light or heavy quark final states are favored.  
Generally, the result is that gluino masses less than about $1$\,TeV and wino masses less than about $300$\,GeV are excluded.
We will also consider cases in which a charged wino in the final state survives into the transition radiation tracker (TRT), and leaves a disappearing charged track.  
Looking for such a disappearing track can allow for a significant reduction in backgrounds, 
while also reducing the event rate.  
We find that this specialized search currently yields weaker limits on the model, 
due to the distance of the TRT from the beam pipe ($\sim 1$m) being much longer than the decay length of the charged wino ($\sim 5$cm).  
These limits could be significantly improved if future searches looked for disappearing tracks in the inner detectors rather than the TRT.   We will also note that in the case that the gluino is heavier than $1$ TeV, the current best limit on the wino mass comes from LEP, and requires $M_2 > 92$ GeV at $95 \%$ confidence.

We will consider future discovery prospects in section 3.2.  We find that after 300 ${\rm fb}^{-1}$ of integrated luminosity at the 14\,TeV LHC, the gluino pair production process considered above, not assuming any charged track information, will be able to probe gluino masses up to about $2.4$\,TeV, and  wino masses up to about $1$\,TeV.  We will also discuss prospects for searches which take advantage of 
the missing charged track signature at the 14 TeV LHC, both with direct wino production, and gluino decays to charged winos.  
We will discuss the event rates for these processes, although limits are difficult to predict without currently knowing the standard model backgrounds for the charged track events.  We will parametrize the rate for such background events, and show how the possible limits might depend on these rates.  We will summarize and conclude in section 4.

\section{Pure gravity mediation}
\label{sec: PGM}

\subsection{Generic features}
Here we will briefly review the pure gravity mediation model. In this model, the only new ingredient other than the MSSM sector is a (dynamical) SUSY breaking sector. Furthermore, the model does not need a singlet SUSY breaking field as required in conventional gravity mediation models. This is quite advantageous because the model is then completely free from the Polonyi problem~\cite{Polonyi}.\footnote{See also reference \cite{hep-ph/0605252} for the Polonyi problem in dynamical SUSY breaking models.} 

Scalar fields in the MSSM sector obtain soft SUSY breaking mass terms by tree-level interactions in supergravity. With a generic K\"ahler potential, all the SUSY breaking masses of the scalar bosons are expected to be of the order of the gravitino mass, $m_{3/2}$~\cite{Nilles:1983ge}. In the following discussions, we will simply assume the following relation,
\begin{eqnarray}
M_{\rm SUSY}^2 \simeq m_{3/2}^2,
\end{eqnarray}
although the details of the spectra do not alter the discussions significantly. Soft SUSY breaking scalar tri-linear couplings are, on the other hand, expected to be suppressed in supergravity at tree-level, so that in this paper we will simply take them to be $0$.

In the PGM model, we further assume that the supersymmetric and SUSY breaking Higgs mixing parameters are generated through tree-level interactions in supergravity (see references \cite{Ibe:2011aa, Ibe:2012hu}). It should be emphasized that those mixing parameters are generated without having a singlet SUSY breaking field~\cite{Inoue:1991rk}. The resultant Higgs mixing parameters, $\mu_H$ and $B_H$, are predicted to be of the order of $m_{3/2}$;
\begin{eqnarray}
\label{eq:HiggsMixing}
 \mu_H = {\cal O}(m_{3/2}), \quad B_H= {\cal O}(m_{3/2}),
\end{eqnarray}
because they are generated via tree-level interactions in supergravity.

Gaugino masses are dominated by one-loop contributions in supergravity, i.e. the anomaly mediated contributions~\cite{Giudice:1998xp, Randall:1998uk,hep-ph/9205227}, as well as additional contributions from threshold effects of the heavy higgsinos~\cite{Giudice:1998xp,Ibe:2006de,  Gherghetta:1999sw}. The gaugino masses are therefore suppressed by loop factors in comparison with those of the scalar bosons and the Higgs mixing parameters.

Altogether, with the requirements that the gauginos (especially, the weak gauginos) are at most in the TeV range so that the lightest neutralino can be a viable candidate for dark matter, the PGM model predicts the following:
\begin{itemize}
\item The squarks, sleptons, and gravitino are similar in mass, with masses in the tens to thousands of TeV range.
\item The higgsinos and the heavier Higgs bosons are also at similar masses.
\item The wino, bino, and gluino are in the hundreds of GeV to tens of TeV range.
\item The Higgs mixing angle is of the order of unity, i.e. $\tan\beta = {\cal O}(1)$.
\end{itemize}
The prediction for the mixing angle $\tan\beta$ results from the fact that all of the scalar masses as well as the higgsino mass parameters are of the order of $m_{3/2}$, i.e.
\begin{eqnarray}
\sin2\beta = 2 B_H \mu_H/m_A^2 = {\cal O}(1),
\end{eqnarray}
where $m_A$ denotes the mass of the heavy Higgs bosons,\footnote{We assume that a linear combination of the Higgs doublet bosons, $h \simeq (\sin \beta) H_u - (\cos\beta) H_d^*$ is light, which is to be obtained via some degree of fine tuning amongst the Higgs mass parameters.} $m_A^2 \equiv m_{H_u}^2 + m_{H_d}^2 + 2|\mu_H|^2$.

\subsection{Details on the gaugino masses}

We obtain the gaugino masses by solving the renormalization group equations with boundary conditions given by the anomaly mediated  and higgsino threshold corrected values at $M_{\rm SUSY} \simeq m_{3/2} \simeq {\cal O}(100)$ TeV. The resultant numerical values are
\begin{eqnarray}
\label{eq:gluino}
m_{\rm gluino} &\simeq&
2.5\times (1 - 0.13 \, \delta_{32} - 0.04 \, \delta_{\rm SUSY})
\times 10^{-2} \, m_{3/2},
\\ \label{eq:wino}
m_{\rm wino} &\simeq&
3.0\times(1 - 0.04 \, \delta_{32} + 0.02 \, \delta_{\rm SUSY})
\times 10^{-3} \, (m_{3/2} + L),
\\ \label{eq:bino}
m_{\rm bino} &\simeq&
9.6\times(1 + 0.01 \, \delta_{\rm SUSY})
\times 10^{-3} \, (m_{3/2} + L/11),
\end{eqnarray}
where $\delta_{\rm SUSY} = \log[M_{\rm SUSY}/100 \,{\rm TeV}]$, $\delta_{32}$ denotes $\delta_{32} = \log[m_{3/2}/100 \, {\rm TeV}]$ for the gluino, and $\delta_{32} = \log[(m_{3/2} + L) /100 \,{\rm TeV}]$ for the wino. The terms proportional to $m_{3/2}$ in the above formulae represent the anomaly mediated contributions, while those proportional to $L$ are the higgsino threshold contributions, where $L$ is defined by
\begin{eqnarray}
L \equiv \mu_H \sin 2\beta
\frac{m_A^2}{|\mu_H|^2 - m_A^2} \ln \frac{|\mu_H|^2}{m_A^2}
\label{eq:L}.
\end{eqnarray}
As discussed in references \cite{Ibe:2011aa, Ibe:2012hu}, $L$ is of the order of the gravitino mass in the PGM model. The wino mass therefore obtains comparable contributions from both the anomaly mediated effects and those of the higgsino threshold corrections.

\begin{figure}[t]
\begin{center}
\includegraphics[width=0.55\linewidth]{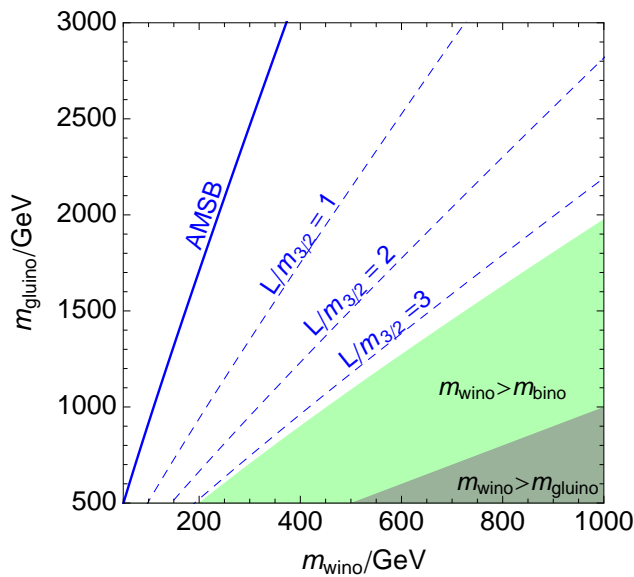}
\caption{\sl \small The correlation between the gluino and wino masses in pure gravity mediation for a given $L$. The solid blue line corresponds to the prediction for purely anomaly mediated gaugino masses, i.e. $L = 0 $. In the light shaded region, the wino mass becomes heavier than the bino mass due to large higgsino threshold effects ($L\gtrsim 3 m_{3/2}$).}
\label{fig:gluinowino}
\end{center}
\end{figure}

In figure \ref{fig:gluinowino}, we show the correlation between the gluino and wino masses from equations (\ref{eq:gluino}) and (\ref{eq:wino}) for a given $L$. The solid blue line shows the prediction for the purely anomaly mediated gaugino masses, i.e. $L=0$, while the dashed lines are for $L/m_{3/2} = 1,2,3$. The gluino mass is about $8$ times heavier than the wino mass for $L=0$, while the ratio gets smaller for positive $L$. In most of the parameter space, the wino is the lightest supersymmetric particle (LSP). A wino LSP with $m_{\rm wino} \lesssim 1000$ GeV is particularly interesting~\cite{ Ibe:2011aa,Ibe:2004tg}, because it can be a viable candidate for dark matter when produced non-thermally by gravitino decays in the early universe, with a reheating temperature appropriate for thermal leptogenesis~\cite{leptogenesis}.\footnote{The thermal relic density of the wino is consistent with the observed dark matter density for the heavier mass, $m_{\rm wino} \simeq 2.7$ TeV~\cite{hep-ph/0610249}, due to a non-perturbative enhancement of the annihilation cross section. For other discussions on non-thermally produced winos, see e.g. \cite{Gherghetta:1999sw, hep-ph/9906527, ArkaniHamed:2006mb}.} It is also worth noting that the PGM model is free from the cosmological gravitino problem, since the gravitino is sufficiently heavy and decays before the start of Big-Bang Nucleosynthesis~\cite{kkm}. In the light shaded region, on the other hand, the wino mass becomes heavier than the bino mass due to large higgsino threshold effects. In this region, the relic density of the bino easily exceeds the observed one due to its highly suppressed annihilation cross section for ${\cal O}(100)$ GeV masses. Thus, the PGM model in this region does not have a viable cosmological scenario. In the following analysis, we concentrate on the parameter space with the wino LSP. In this parameter space, the figure shows that the gluino can be as light as twice the wino mass, which makes the PGM model more accessible than in a conventional anomaly mediation model at the LHC experiment.

For later convenience, we discuss detailed properties of the wino which are relevant for the LHC physics. As an important feature, its neutral component (the wino LSP $\tilde \chi^0$) is almost degenerate with its charged one (the charged wino $\tilde \chi^\pm$) due to approximate custodial symmetry. The dominant mass splitting between them comes from one-loop gauge boson contributions~\cite{Feng:1999fu}, and the resultant splitting is
\begin{eqnarray}
\label{eq:deltaM}
{\mit \Delta} m_{\rm wino}  = m_{\tilde\chi^\pm}- m_{\tilde\chi^0} =
\frac{g_2^2}{16\pi^2} m_{\rm wino}
\left[ f(r_W) - \cos^2 \theta_W f(r_Z) - \sin^2 \theta_W f(0) \right],
\end{eqnarray}
where $f(r)= \int^1_0 dx (2 + 2 x^2) \ln[x^2 + (1 - x)r^2]$ and $r_{W,Z}=m_{W,Z}/m_{\rm wino}$. For a wino mass in the hundreds of GeV range, the splitting is ${\mit \Delta} m_{\rm wino} \simeq 160-170$ MeV, and hence, the neutral and the charged wino are highly degenerate. Due to this fact, the dominant decay mode of the charged wino is into a neutralino and a charged pion, ${\tilde \chi}^\pm \to {\tilde \chi}^0 + \pi^\pm$. The decay rate of this mode is given by
\begin{eqnarray}
\Gamma({\tilde \chi}^\pm \to {\tilde \chi}^0 + \pi^{\pm}) = 
\frac{2 G_{F}^{2}}{\pi} \cos^2 \theta_c f_\pi^2
{\mit \Delta} m_{\rm wino}^3
\left(1- \frac{m_\pi^2}{{\mit\Delta}m_{\rm wino}^2}\right)^{1/2},
\label{eq:WinoWidth}
\end{eqnarray}
where $G_F \simeq 1.17 \times 10^{-5} \,{\rm GeV}^{-2}$, $f_\pi \simeq 130$ MeV, and $\theta_{c}$ is the Cabbibo angle. The charged wino has therefore a rather long lifetime for the purpose of collider physics,
\begin{eqnarray}
\tau_{\rm wino} \simeq 1.4 \times 10^{-10} {\rm sec}
\left( \frac{160 \, \rm MeV}{\mit \Delta m_{\rm wino}} \right)^3
\left(1 - \frac{m_\pi^2}{{\mit\Delta}m_{\rm wino}^2} \right)^{-1/2},
\end{eqnarray}
and the charged wino produced at the LHC experiment travels typically ${\cal O}(1-10)$ cm before it decays. In the following sections, we will scrutinize how this property could potentially enhance the detectability of the gauginos at the LHC.

\subsection{The lightest Higgs boson mass}

Before closing this section, we would like to highlight how a light Higgs boson mass, as reported very recently by the ATLAS and CMS collaborations, can be consistently explained within the framework of the PGM model. As we have summarized above, the model with the gauginos in the TeV range forces the scalar masses to be in the hundreds of TeV range. In this case, the mass of the lightest Higgs boson is inevitably heavier than that predicted in conventional MSSM models thanks to large renormalization group effects on the Higgs quartic coupling~\cite{TU-363}.

\begin{figure}[t]
\begin{center}
\includegraphics[width=0.55\linewidth]{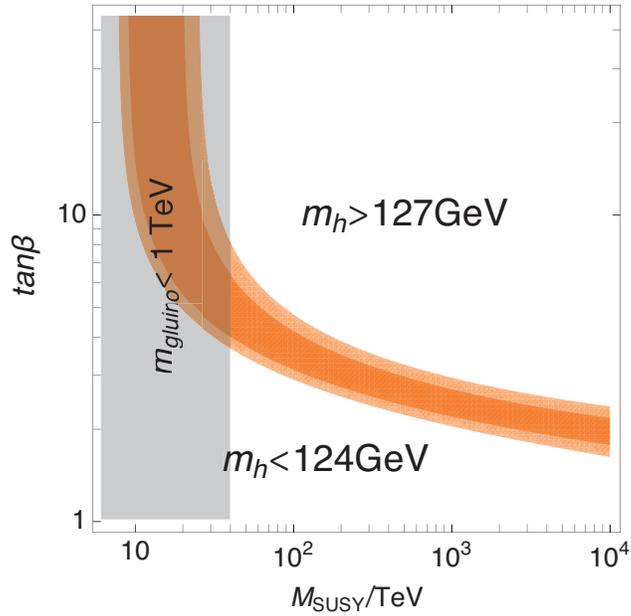}
\caption{\sl \small The contour plot of the lightest Higgs boson mass. Here, we have fixed $m_{3/2} = \mu_H = M_{\rm SUSY}$. The orange band shows the Higgs boson mass 124 GeV $< m_h < 127$ GeV observed by the ATLAS and CMS collaborations for the central value of the top quark mass. The light orange band is the one including the $1\sigma$ error of the top quark mass. In the gray shaded region, the gluino mass is below about 1 TeV, which is predicted by the anomaly mediation in equation (\ref{eq:gluino}) for $M_{\rm SUSY} = m_{3/2}$. The detailed discussion on the constraint on the gluino mass is discussed in the following section.}
\label{fig:Higgs}
\end{center}
\end{figure}

In figure \ref{fig:Higgs}, a contour plot of the Higgs boson mass is shown as a function of $M_{\rm SUSY}$ and $\tan\beta$. In order to obtain the Higgs boson mass, we have solved the one-loop renormalization-group equations in accordance with references \cite{hep-ph/0406088, arXiv:0705.1496, arXiv:1108.6077} with a boundary condition given by the so-called SUSY relation,
\begin{eqnarray}
\label{eq:SUSY}
\lambda = \frac{1}{4} \left(\frac{3}{5}g_1^2 + g_2^2 \right) \cos^2 2\beta,
\end{eqnarray}
at the energy scale of $M_{\rm SUSY}$. Threshold corrections at the heavy scalar scale are also included. It is worth noting that the predicted Higgs boson mass is slightly lighter than the one in reference \cite{arXiv:1108.6077}, because, unlike in Split SUSY models~\cite{hep-ph/0406088, hep-th/0405159,  hep-ph/0409232}, $\mu_H$ here is in the range of the gravitino mass.

The orange band shows Higgs boson masses between 124 GeV--127 GeV, as supported by the LHC experiments, with the central value of the top quark mass in $m_{\rm top} = 173.18 \pm 0.94$ GeV~\cite{Aaltonen:2012ra}. The light orange band comes from including the $1\sigma$ error on the top quark mass. In the figure, the gray shaded region corresponds to $m_{\rm gluino} < 1$ TeV as predicted by pure anomaly mediation according to equation (\ref{eq:gluino}) with $M_{\rm SUSY} = m_{3/2}$. As we will discuss in the following section, gluino masses below about 1 TeV have been excluded by the LHC experiments. Thus, the figure shows that the model requires a rather small $\tan\beta = {\cal O}(1)$ to  explain the observed Higgs boson mass with 124 GeV $\lesssim m_h \lesssim$ 127 GeV consistently. As we emphasized above, the size of $\tan\beta$ in this model is predicted to be of a roughly appropriate size due to the large Higgs mixing parameters.

\section{LHC signals of the PGM model}
\label{sec: LHC signals}

We are now in a position to discuss the collider signals of the model. The PGM model predicts the following three signals at the LHC experiment. The first signal is the pair production of the gluino. Once the gluino is produced, it decays mainly into two quarks and a charged/neutral wino unless right-handed squarks are much lighter than left-handed squarks.\footnote{In such a case, the gluino decays mainly into two quarks and a bino, and the bino decays into a charged wino by emitting a $W$ boson or into a neutral wino by emitting a Higgs boson. It is also worth noting that, for a higgsino mass parameter $\mu_H$ of the order of 10--100 TeV, the radiative decay of the gluino into a gluon and a neutralino is very suppressed, unlike in the case of conventional anomaly mediated SUSY breaking or Split SUSY models~\cite{gluino radiative decay}.} The charged wino eventually decays into a neutral wino by emitting a soft pion, though the pion is hardly detected. The second signal is again from the pair production of the gluino, but we focus only on the gluino decaying into a charged wino. Since the decay length of the charged wino (its lifetime times the speed of light) is about 5 cm, it sometimes leaves the signal of a disappearing charged track in the inner detector. The existence of the disappearing track is one of the distinct signals of the PGM model and it can be used to reduce standard model (SM) backgrounds significantly. The last signal is the direct production of the charged wino through electroweak interactions. Though the cross section of the direct production is not large, the possibility of a disappearing track  can enable us to find a signal of the charged wino despite SM backgrounds.

In the following, we first consider the current bounds on gluino and wino masses obtained with 5 fb$^{-1}$ of data from the LHC experiment at $\sqrt{s} =$ 7 TeV running. We put the bounds in each process mentioned above, where the processes are referred to as 
\begin{itemize}
\item Gluino pair production without use of the disappearing track information,
\item Gluino pair production with use of the disappearing track information,
\item Direct wino production with use of the disappearing track information.
\end{itemize}
After determining the bounds, we discuss the capability of the LHC experiments to find collider signals of the PGM model in the future. We investigate the gluino and neutralino mass reach assuming 300 fb$^{-1}$ data at 14 TeV running.

\subsection{Current bounds}

\begin{figure}[t]
\begin{center}
\includegraphics[width=0.65\linewidth]{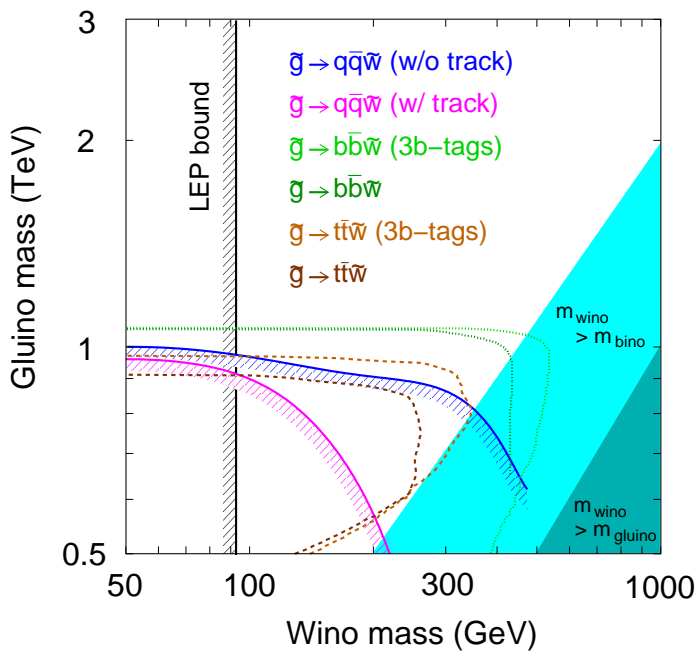}
\caption{\sl \small Bounds on the gluino and wino masses in the PGM model obtained with 5 fb$^{-1}$ of data at 7 TeV running. The solid blue line ($\tilde{g} \to q \bar{q} \tilde{w}$ (w/o track)) is obtained via the gluino pair production without use of the disappearing track information, while the solid magenta line ($\tilde{g} \to q \bar{q} \tilde{w}$ (w/ track)) is obtained with use of the disappearing track information. The LEP bound on the wino mass is also shown (from the radiative return process). For other lines, please see the text. The region with $m_{\rm bino} < m_{\rm wino}$ is shaded because this region is not favored from the viewpoint of cosmology (overclosure of the universe).}
\label{fig: constraints}
\end{center}
\end{figure}

We first consider the bounds obtained from gluino pair production without use of the disappearing track information. The event topology of the signal is multiple jets with large missing momentum, which is one of the typical SUSY signals and studied well by both the ATLAS and CMS collaborations. Based on the strategy of object reconstructions and kinematical selections from reference \cite{ATLAS-CONF-2012-033}, we have calculated the cross section (times acceptance) of the signal process using the Pythia~\cite{Sjostrand:2006za} and Delphes~\cite{Ovyn:2009tx} codes.\footnote{Our simulation framework has been verified by reproducing event distributions and corresponding cut-flow table of the signal process presented in reference \cite{ATLAS-CONF-2012-033} using appropriate sample points. The cross section has been calculated at NLO using the Prospino~\cite{Beenakker:1996ed} code in this calculation.} We have assumed that all gluinos decay into two quarks (except the top quark) and a neutral/charged wino. Results of the calculation for various gluino and wino masses were compared to the experimental upper limit shown in reference \cite{ATLAS-CONF-2012-033}, and we obtained the bounds on gluino and wino masses at 95\% confidence level, as shown in figure \ref{fig: constraints} by the solid blue line. It turns out that the region with $m_{\rm gluino} \lesssim$ 1 TeV and $m_{\rm wino} \lesssim$ 300 GeV has been excluded.

When the gluino decays into bottom quarks rather than light quarks, the bounds are expected to be stronger, because the tagging of b-jets can reduce SM backgrounds efficiently. On the other hand, when the gluino decays mainly into top quarks, the bounds will be weaker, because the signal event contains many jets which make the analysis complicated. In figure \ref{fig: constraints}, as a reference, we also depict the bounds  assuming that all gluinos decay into two bottom quarks as a dotted dark-green line~\cite{gluino decay in CMS}, or into two top quarks as a dashed dark-brown line~\cite{gluino decay in ATLAS}. The bounds become stronger (weaker) as expected when the gluino decays into $b\bar{b}$ ($t\bar{t}$). Very recently, a new limit on the gluino pair production followed by the gluino-decay into two bottom quarks or two top quarks has been reported by tagging more than two b-jets~\cite{ATLAS-CONF-2012-058}. The bounds have been improved as can be seen in the figure (the dotted light-green line and  dashed light-brown line). Since the pattern of the gluino decay depends on the details of the squark mass spectrum, and the bounds on the gluino and wino masses are not significantly different in either case, in the following discussions we simply assume that all gluinos decay into two light quarks and a neutral/charged wino as a representative example.  We also assume to be explicit that one-third of gluinos decay into light quarks by emitting a neutral wino and the other two-thirds by emitting a charged wino.

We next consider the bounds from gluino pair production with use of the disappearing track information. We have utilized a result presented in reference \cite{disappearing track}, where model-independent upper limits on the cross section (times the acceptance) for non-SM physics events with an isolated disappearing track are provided. Based on object reconstructions and kinematical selections adopted in this reference, we have calculated the cross section in the PGM model using the Pythia, Prospino, and Delphes codes. Since the code cannot deal with information of the disappearing charged track very accurately, we have imposed the following selection criteria on generated events in order to obtain a disappearing charged track which is sufficiently isolated from other objects.
\begin{itemize}
\item
The event must have at least one high $p_T$ isolated track within $|\eta|<$0.63, where $p_T$ and $\eta$ are the transverse momentum and pseudo-rapidity of the track.
\item
The track is regarded as isolated if there is no track with $p_T>$ 0.5 GeV within a distance $\Delta R$= 0.1, where $\Delta R = (\Delta \eta ^2 + \Delta \phi^2)^{1/2}$ with $\Delta \eta$ and $\Delta \phi$ being the pseudo-rapidity and azimuthal angle between the two tracks.
\item
The selected track must disappear between 514 mm and 863 mm, i.e. within the first and second layers of the transition radiation tracker (TRT).
\end{itemize}
We have confirmed that the simulation framework mentioned above reproduces the cut-flow table presented in reference \cite{disappearing track} in which several representative points of the conventional anomaly-mediation model were discussed. The resulting bounds on the gluino and wino masses are shown in figure \ref{fig: constraints} as a solid magenta line. Since the location of the TRT is 1 m away from the beam pipe and the typical decay length of the charged wino is 5 cm, the bounds are weaker than those obtained by the conventional analysis (the gluino pair production without use of the disappearing track information). Note that, in particular,  when the wino mass becomes heavier, the gamma-factor of the charged wino becomes smaller, which results in weaker limits on gluino production. If more inner detectors were used  to look for disappearing tracks, on the other hand, the bounds would become much stronger, as will be discussed in the  next subsection.

Finally we consider the bound on the wino mass which is obtained from the direct wino production with use of the disappearing track information. Though the current LHC data potentially can exclude a wino lighter than about 100 GeV using the wino pair production associated with a jet~\cite{direct wino constraint}, unfortunately no official bound has been reported yet. The only bound we can place at this time comes from the LEP experiment, and as such the wino mass is constrained to be heavier than 92 GeV at 95\% confidence level~\cite{LEP bound}. This constraint has been obtained by searching for the radiative return process from wino pair production ($e^+ e^- \to \gamma \tilde{\chi} \tilde{\chi}$ with $\tilde{\chi}$ being a charged or a neutral wino). This LEP bound is also shown in figure \ref{fig: constraints} as a solid black line.

\subsection{Future prospects}

\begin{figure}[t]
\begin{center}
\includegraphics[width=0.65\linewidth]{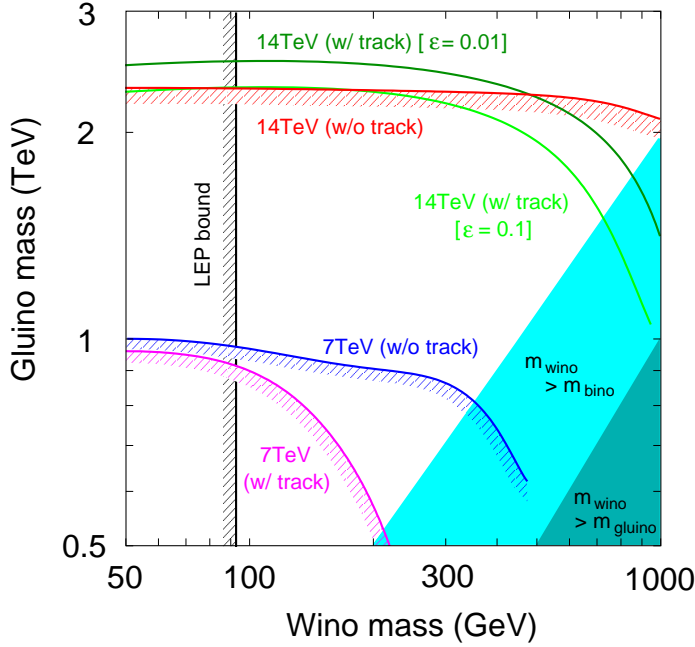}
\caption{\sl \small Expected bounds on the gluino and wino masses obtained with 300 fb$^{-1}$ of data at 14 TeV running. The solid red line (14 TeV (w/o track)) is obtained by the gluino pair production without use of the disappearing track information, while the solid dark-green line (14 TeV (w/ track) [$\epsilon = 0.01$]) and light-green line (14 TeV (w/ track) [$\epsilon = 0.1$]) are obtained by the production with use of the disappearing track information for the cases of $\epsilon =$ 0.01 and 0.1, respectively. Other lines are the same as those in figure \ref{fig: constraints}. In all cases, the gluino is assumed to decay into two light quarks and a charged/neutral wino.}
\label{fig: prospects}
\end{center}
\end{figure}

As in the discussion in the previous subsection, we first consider the expected bounds on the gluino and wino masses obtained from the gluino pair production without use of the disappearing track information. We have assumed an accumulation of 300 fb$^{-1}$ of data at the 14 TeV LHC, and utilized the result in reference~\cite{TestAMSB} to estimate the SM backgrounds. Based on the object reconstructions and kinematical selections adopted in this reference, we have calculated the cross section (times acceptance) of the signal process for various gluino and wino masses using the Pythia, Prospino, and Delphes codes.\footnote{Our simulation framework has been verified by reproducing the event distribution of the signal process as a function of the effective mass with use of the sample point adopted in reference~\cite{TestAMSB}.} These results have been compared with the SM backgrounds in the reference and we then obtained the expected bounds at 95\% confidence level, which are shown in figure \ref{fig: prospects} as the solid red line. The region with $m_{\rm gluino} \lesssim 2.4$\,TeV and $m_{\rm wino} \lesssim 1$\,TeV will be covered at the future LHC experiment.

We next consider the expected bounds on the gluino and wino masses from  gluino pair production with the use of the disappearing track information. We have calculated the cross section (times acceptance) of the signal process using the Pythia, Prospino, and Delphes codes again. Object reconstructions and kinematical selections are the same as above- however, in order to obtain an isolated and charged disappearing track, the following selection criteria were also imposed on each generated event:
\begin{itemize}
\item
The event must have at least one high $p_T$ isolated track within $|\eta|<$1.4.
\item
There must be no other tracks with $p_T>$ 2.0 GeV within $\Delta R$= 0.1 of the isolated track.
\item
The selected track must disappear between 142 mm and 520 mm, i.e. between the inner pixel detectors and the semiconductor detector (SCT).
\end{itemize}
SM backgrounds against this signal are, on the other hand, difficult to estimate without the use of real data. We have therefore introduced a parameter $\epsilon$ in order to describe how efficiently the SM backgrounds can be reduced when we impose the selection criteria for the isolated and charged disappearing track. The cross section of the SM backgrounds is then estimated as $\epsilon$ times $\sigma_{\rm BG}$ with $\sigma_{\rm BG}$ being the background cross section (times acceptance due to other selection criteria) without imposing the existence of the track. Results for the signal cross section for various gluino and wino masses have been compared with the resulting SM backgrounds to obtain the expected bounds at 95\% confidence level, shown in figure \ref{fig: prospects} as the solid dark-green and light-green lines for the cases of $\epsilon = 0.01$ and $0.1$, respectively. When the wino is relatively light and $\epsilon$ is small enough, this process gives stronger bounds than were obtained without use of the track. In our analysis, we have assumed somewhat optimistic parameters for finding the track- namely, we are allowing the charged winos to decay between 142 mm and 520 mm. If it is in fact necessary to use more outer detectors to find the track, then the bounds will be weaker as discussed in reference \cite{Kane:2012aa}. On the other hand, if the SM backgrounds can be reduced more efficiently- namely, if smaller $\epsilon$ can be used- then the bounds will be stronger than those presented here.

Finally we consider direct production of the wino through electroweak interactions at the future LHC experiment (300 fb$^{-1}$ \& 14 TeV). Since the cross section of the signal process is much smaller than that for gluino production, it is difficult to determine the expected bound on the wino mass without having accurate data for the SM backgrounds. We therefore briefly discuss some prospects of this process for constraining the PGM model using the signal cross section at parton level.

\begin{figure}[t]
\begin{center}
\includegraphics[width=0.65\linewidth]{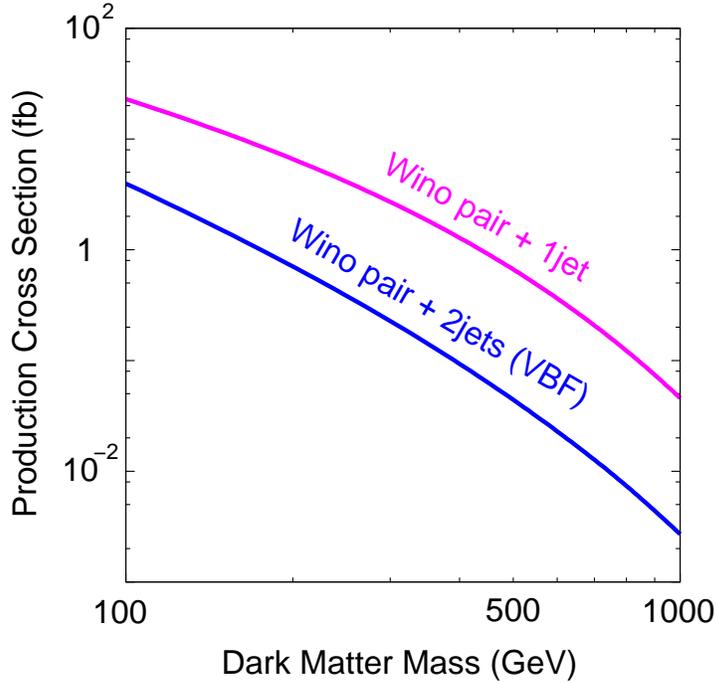}
\caption{\sl \small Cross sections for the direct wino production at the 14 TeV LHC with 300 fb$^{-1}$ of data. The solid magenta line (wino pair + 1 jet) is the wino pair production associated with a jet (a colored parton), while the solid blue line (wino pair + 2 jets (VBF)) is the wino pair production through vector boson fusion processes. These cross sections are obtained after applying appropriate selection cuts. For details, please see the text.}
\label{fig: DirectWino}
\end{center}
\end{figure}

There are actually two production processes. One is the pair production of the wino associated with a jet~\cite{Ibe:2006de, Moroi:2011ab}. Based on the kinematical cuts in reference~\cite{Beltran:2010ww} to reduce QCD and electroweak backgrounds, we have calculated the cross section for wino production associated with a colored parton using the CalcHep code~\cite{CalcHep}, which is shown in figure \ref{fig: DirectWino} as a solid magenta line. We also imposed the following selection cuts in order to have a charged track in the central region of the detector:
\begin{itemize}
\item The final state of the process should involve at least one charged wino.
\item The charged wino should be within the central region, namely, $|\eta| < 1.7$.
\end{itemize}
Since the SU(2)$_L$ charge of the wino is one, the cross section of the signal process exceeds 1 fb unless the wino mass is too heavy. On the other hand, the cross section of the SM backgrounds after applying the kinematical cuts (but before applying the selection cuts for the charged track) is about 700 fb. 

A second process is the pair production of the wino through vector boson fusion (VBF). Based on the kinematical cuts in reference~\cite{VBF wino}, which requires the existence of two forward jets (colored partons) with a large rapidity gap and a suppression of jet activities in the central region of the detector, we have calculated the cross section of the signal process using the CalcHep code.\footnote{Reference \cite{VBF wino} analyzed the invisible decay of the Higgs boson through VBF processes. We have therefore relaxed the kinematical cut on the invariant mass between two forward jets (two forward colored partons) as $M_{jj} >$ 800 GeV, because the missing transverse energy is increased thanks to a pair of winos in signal events and this reduces the SM backgrounds.} We have also considered the same selection cuts for the charged wino mentioned above in this calculation. The result is shown in figure \ref{fig: DirectWino} as a solid blue line. The cross section for the VBF process is 5--10 times smaller than that for the wino pair production in association with a jet. The cross section of the SM backgrounds on the other hand is about 170 fb, which we note is much smaller than in the case of associated production. One of the attractive features of the use of the VBF process is that, since this process provides a clean environment at the central region of the detector, the charged disappearing track will be efficiently searched for, though the absolute number of signal events is small.

The direct production of the wino is of importance to investigate the PGM model because it gives a bound on the wino mass independent of the gluino mass. Finding more sophisticated kinematical selections to increase the number of signal events and also, use of the most inner detectors, are therefore both ardently desired.

\section{Summary}
\label{sec: summary}

Pure gravity mediation of supersymmetry breaking is a very attractive and simple scenario for physics beyond the SM.   It predicts squarks, sleptons, higgsinos, heavy Higgs bosons, and the gravitino to be of the order of 10 to 1000 TeV, while the gluino, bino, and wino all remain in the roughly TeV range. As a result, it does not have any supersymmetric flavor/CP problems, and the Higgs mass is predicted to be about 125 GeV, consistent with recent results reported by the ATLAS and CMS collaborations. It is also worth noting that the fact that all squarks are very heavy in this scenario is consistent with the so-far negative results of new physics searches at the LHC.

The PGM model is attractive also from the viewpoint of cosmology. Since the model does not need any singlet fields in the SUSY breaking sector, we do not have to worry about the so-called the Polonyi problem. The heavy gravitino mass is also beneficial for the generation of the baryon asymmetry of the universe, because the model is compatible with the thermal leptogenesis scenario thanks to the decay of the gravitino before BBN. In addition, the model predicts an attractive candidate for dark matter, the neutral wino. Since its annihilation cross section is boosted by the Sommerfeld enhancement~\cite{Sommerfeld}, we may be able to find strong signals in various indirect detection measurements~\cite{Ibe:2012hu}.

In this article, we have  focused mainly on collider signals of the pure gravity mediation model at the LHC experiment. The following three processes were found to be important as potential signals:
\begin{itemize}
\item[(a)] Gluino pair production without the use of disappearing wino charged track information.
\item[(b)] Gluino pair production with the  use of disappearing track information.
\item[(c)] Direct wino production with the use of disappearing track information.
\end{itemize}
We have found that process (a) gives us the most severe constraints on the gluino and wino masses at the current stage of the LHC experiment (7 TeV \& 5 fb$^{-1}$). The region with $m_{\rm gluino} \lesssim$ 1 TeV and $m_{\rm wino} \lesssim$ 300 GeV has been excluded. Process (b) also gives some constraints on the masses. Since the TRT detector which is far away from the beam pipe has been used to search for disappearing tracks, these constraints are weaker than those obtained from process (a), since most charged winos decay too early. So far no official results on process (c) have been reported yet, but the LEP experiment excludes the region with $m_{\rm wino} <$ 92 GeV.

We have  also considered the prospects  for each of these signals at the future LHC experiment (14 TeV \& 300 fb$^{-1}$). We have found that process (a) will cover the region with $m_{\rm gluino} \lesssim 2.4$\,TeV and $m_{\rm wino} \lesssim 1$\,TeV. Process (b) will play an important role to search for pure gravity mediation signals if the most inner detectors can be used to search for disappearing tracks. Process (c) is also important because it will allow us to search for the wino independent of the gluino mass. We have found in particular that wino pair production through vector boson fusion is interesting, because it provides a clean environment in the central region to aid in finding the track. Since the cross section for process (c) is small, it will be important to try to find more sophisticated kinematical selections in order to enhance the number of signal events.

\section*{Acknowledgments}

This work is supported by the Grant-in-Aid for Scientific research from the Ministry of Education, Science, Sports, and Culture (MEXT), Japan, No. 24740151 (M.I.), No. 22244021 (S.M. and T.T.Y.), No. 23740169 (S.M.), and also by the World Premier International Research Center Initiative (WPI Initiative), MEXT, Japan.


\end{document}